\documentclass{article}

\usepackage{arxiv} %
\usepackage{amsmath,amssymb,amsfonts}
\usepackage{graphicx,subfig}
\usepackage{xspace}
\usepackage{proof}
\usepackage[colorlinks,allcolors=blue]{hyperref}

\let\vec\mathbf
\let\mat\vec

\newcommand\dLrule[1]{\textbf{\textcolor{blue}{#1}}}
\newcommand\norm[1]{\lVert#1\rVert}
\newcommand\hpr[1]{\mathsf{#1}}
\newcommand\dbox[1]{[#1]}
\newcommand\ddia[1]{\langle#1\rangle}

\makeatletter
\begin{document}

\title{
  How Deduction Systems Can Help You To Verify Stability Properties
	}

	\author{Mario Gleirscher$^1$, Rehab Massoud$^3$, Dieter 
Hutter$^{1,2,*}$, Christoph Lüth$^{1,2,*}$ 
  \thanks{$^{1}$ University of Bremen, 28359 Bremen, Germany.}%
  \thanks{$^{2}$ Cyber-Physical Systems, DFKI, 28359 Bremen, Germany.}%
  \thanks{$^{3}$ EPFL,Systems Integration Lab, 1015 Lausanne, Switzerland. Most 
of the work was done before joining EPFL, while affiliated with $^{2}$.}%
  \thanks{$^{*}$ Supported by the VeryHuman project (grant number 01IW20004) 
funded the German Federal Ministry of Education and Research (BMBF).}%
  }%
  
\maketitle
	
\begin{abstract}
Mathematical proofs are a cornerstone of control theory, and it is important to 
get them right. Deduction systems can help with this by mechanically checking 
the proofs. However, the structure and level of detail at which a proof is 
represented in a deduction system differ significantly from a proof read and 
written by mathematicians and engineers, hampering understanding and adoption 
of these systems.

This paper aims at helping to bridge the gap between machine-checked proofs and 
proofs in engineering and mathematics by presenting a machine-checked proof for 
stability using Lyapunov's theorem in a human-readable way. The structure of 
the proof is analyzed in detail, and potential benefits of such a proof are 
discussed, such as generalizability, reusability and increased trust in 
correctness.
\end{abstract}

\section{Introduction}
	
Stability assurance is essential for the safe and reliable performance of 
controlled systems, such as autonomous robots or vehicles.  A prominent 
safety-related issue is to guarantee that a controller combined with a machine 
constitutes a stable controlled system.  The correct behavior of the controller 
has to be guaranteed rigorously in that context.   In many cases, Lyapunov's 
work on stability theory provides the mathematical foundation for the required 
verification, delivering a framework for organizing corresponding proofs.

  Often, such proofs are tedious, error-prone, and rely on hidden assumptions, 
thereby creating the desire for a higher automation in finding and checking 
such proofs. Deduction systems (particularly, interactive tools called proof 
assistants) have a long history and are successfully applied in various 
domains, for example, for verifying chip designs~\cite{10.1145/1592434.1592437} 
and for solving configuration or optimization 
problems~\cite{DBLP:conf/hybrid/LoosRP13, 
dawson2023safe,DBLP:conf/rv/MitschP14, DBLP:journals/ral/BohrerTMSP19, 
desai2017combining}. 
	As opposed to mathematicians arguing on a semantic level, deduction 
systems operate on a pure syntactical level applying a fixed set of rules to 
rewrite a given problem until it is subsumed by the given facts. Such a 
syntactic or fully formalized proof can be mechanically checked. Depending on 
the expressiveness of the underlying logic, deduction systems are able to find 
complex proofs automatically or they might need help in particular situations, 
for instance to select	an appropriate application of some proof rule from a 
number of applicable alternatives, or to stop the system from pursuing 
irrelevant paths during proof search.

  In the past decade, deduction systems have significantly 
improved~\cite{Gleirscher2023-ManifestoApplicableFormal}, now able to tackle 
problems in control theory and engineering, including stability assurance. For 
instance, the KeYmaera~X tool~\cite{DBLP:conf/cade/FultonMQVP15} provides 
specific proof support for applying Lyapunov's theorems 
\cite{DBLP:conf/tacas/TanP21} and for verifying the stability of switched 
systems~\cite{DBLP:conf/hybrid/TanMP22}.  For another example, the Coq 
prover~\cite{bertot2013interactive} was used to prove La~Salle's theorem 
\cite{cohen2017formal} and the stability of the inverted pendulum on a 
cart~\cite{rouhling2018formal}. Using such tools to formalize stability proofs, 
one not only obtains more detailed rigorous proofs, but also gains more insight 
and understanding of the problem and its dependencies on the various 
parameters. For example, in~\cite{rouhling2018formal}, authors found and fixed 
errors in a previous manual stability proof~\cite{lozano2000stabilization}. 

  However, using deduction systems for non-trivial applications typically 
requires human interaction during the proof search. Even if the proof is 
well-understood at the mathematical level, it is still a non-trivial task to 
find its formal, 
	syntactically derived counterpart.  Vice versa, it is non-trivial to 
interpret a syntactical proof in its semantic meaning, which is a prerequisite 
for assisting the deduction systems in the next steps of a proof attempt. 
	
	This paper helps bridging this gap between semantic, human-oriented and 
syntactic, formal proofs. We discuss a formal proof of the stability of a 
controlled system using Lyapunov's stability theorem, using the inverted 
pendulum as a running example. Our specific contributions are:	

	\begin{itemize}
		\item We extract the structure (e.g.\ key steps) of a 
mechanization of a fundamental control-theoretic 
proof~\cite{DBLP:conf/tacas/TanP21} and present it in terms familiar to control 
theorists.
		\item We connect a usual approach for problem family 
characterization (i.e.\ identifying healthy combinations of system dynamics and 
Lyapunov function templates) with a deductive proof using well-formedness 
constraints on the parameters of the dynamics and the Lyapunov function 
template.
		\item We replicate and enhance the proof in 
\cite{DBLP:conf/tacas/TanP21} using these constraints as an additional 
side-condition, allowing both, the deductive system and its user, to navigate 
the proof more easily and intuitively.		
	\end{itemize}
		
	The paper is organized as follows: 
Section~\ref{sec:formal-preliminaries} presents the formal background and the 
running example,  before we apply Lyapunov's direct method in 
Section~\ref{sec:stability} to characterise stable pendula.  
Section~\ref{sec:deductive-proof} explains the deductive proof. We then 
highlight in Section~\ref{sec:discussion} the benefits of our approach, before 
we conclude the paper in Section~\ref{sec:conclusion}.

\section{Formal Preliminaries}
\label{sec:formal-preliminaries}

After defining Lyapunov's stability theorem and giving an overview of
differential dynamic logic---the deduction system used in this work for
constructing stability proofs---we introduce the inverted pendulum as
a running example.
Scalars $x\in\mathbb{R}$ are in italic and vectors $\vec x\in\mathbb{R}^n$ in
bold.

\subsection{Lyapunov Stability Theory}
\label{sec:lst}

Given system dynamics $\dot{\vec x}$ and, without loss of generality,
an isolated equilibrium point $\vec x_e=\vec 0\in\mathbb{R}^n$, asymptotic
stability \cite{Khalil2015-Nonlinearcontrol} of $\vec x_e$ amounts to
establishing
\begin{align*}
  \forall\epsilon > 0\,\exists\delta\colon
  \norm{\vec x_0} < \delta
  &\Rightarrow
    \forall t \geq 0\colon \norm{\xi(t;\vec x_0)} < \epsilon
    \quad\text{and}
  \\
  \exists\delta > 0\colon
  \norm{\vec x_0} < \delta
  &\Rightarrow
  \lim_{t\to\infty}\norm{\xi(t;\vec x_0)} = 0
\end{align*}
of trajectories $\xi$ emanating from $\vec x_0$ in a $\delta$
environment.  \emph{Lyapunov's direct method} simplifies this approach
to proving for $\dot{\vec x}$ and $\vec x\neq\vec x_e$ the existence
of a function $V$ with
\begin{align}
  \label{constr:lst-nd}
  V(\vec x)&>0 \quad\text{(positive definiteness of $V$) and}
  \\
  \label{constr:lst-pd}
  \dot V(\vec x)&<0 \quad\text{(negative definiteness of~$\dot V$)},
\end{align}
where $\dot V\equiv\nabla V\cdot\dot{\vec x}$ is the total derivative
of $V$ along $\dot{\vec x}$.  One challenge with the direct method is
finding such a~$V$.

\subsection{Sequent Calculus and Differential Dynamic Logic}
\label{sec:dl-intro}

Sequent calculus is a deduction system going back to Gentzen and Hilbert. 
Basically, a sequent is given as $A_1,\ldots,A_n \vdash B_1,\ldots,B_m$ where 
$A_i$ are the assumptions and $B_j$ the conclusions, meaning that if all $A_i$ 
hold, one of $B_j$ will hold (where $A_i, B_j$ are formulae). Formal proofs are 
represented as proofs trees, labeled at the nodes with sequents, and 
constructed using a handful of inference rules such as the \emph{Cut} rule
\begin{align}
\label{eq:cut}
    \infer[\dLrule{Cut}]{\Gamma,\Sigma \vdash \Pi,\Delta}
          {\Gamma\vdash \Pi, A & A,\Sigma\vdash \Delta}
\end{align}
which allows us to drop (or cut) the formula $A$ from the sequent. Other rules 
allow the manipulation of sequents (adding axioms, dropping assumptions) or 
formulae. Some rules can be applied schematically (such as rules introducing a 
connective), but in particular using the cut rules requires a semantic 
understanding of the proof.

We use \emph{differential dynamic logic} ($\mathsf{d}\mathcal{L}$){} to express properties and 
(hybrid) programs. $\mathsf{d}\mathcal{L}${} allows writing down hybrid programs and properties in 
one language. Programs combine discrete and continuous state transitions; in 
this paper, we are only interested in the latter. A program is given by the 
usual constructs for assignments, choice and iteration, and differential 
equations describing the evolution of the variables. Properties are given by 
first-order formulae of real arithmetic, and the modal operators $\ddia{\hpr P} 
\alpha$ and $\dbox {\hpr P} \alpha$, where $\hpr P$ is a hybrid program and 
$\alpha$ a formula; $\dbox{\hpr P} \alpha$ that means that $\alpha$ always 
holds when and if $\hpr P$ terminates, and $\ddia{\hpr P}  \alpha$ means that 
$\hpr P$ will terminate in at least one state where $\alpha$ is true.

There are rules allowing us to manipulate terms according to the usual rules of 
first-order logic (quantifier elimination/introduction etc.\xspace), and rules for the 
modal operators, such as the \textit{differential Cut} which illustrates the 
interplay between structural logic rules and differential equations:
\begin{align*}
\infer[\dLrule{dC}]
  {\Gamma\vdash\dbox{\dot{\vec x}= f(\vec x)\& Q} P,\Delta}
  {\Gamma\vdash\dbox{\dot{\vec x}= f(\vec x)\& Q}C,\Delta
  & \Gamma\vdash\dbox{\dot{\vec x}= f(\vec x) \& Q \land C} P,\Delta}
\end{align*}
Here, $\dot{\vec x}= f(\vec x) \& Q$ describes a program where $\vec x$ evolves 
as specified as long as $Q$ holds; it terminates once $Q$ does not hold. If we 
can show that $C$ holds whenever the program terminates, and if we can further 
show $P$ with $C$ in the assumptions, we may drop $C$ entirely. This rule is 
similar to the traditional cut rule (\ref{eq:cut}) in that it allows to drop 
assumptions, but not derivable since it also deals with the evolution of $\vec 
x$ over time.

\paragraph*{Auxiliary Notation}
\label{sec:notation}

Given some origin $\vec x_e$, a ball $B_p$ is a closed set
$\{\vec x\in\mathbb{R}^n\mid \norm{\vec x-\vec x_e}\leq p\}$.  With
$\bar B_p$, $B_p^\circ$, and $\partial B_p$, we denote $B_p$'s
complement, interior, and boundary, respectively.  For
$p,q\in\mathbb{R}_{>0}$, $_p\ddia{\hpr P}_q$ and $_p\dbox{\hpr P}_q$ denote
that if program $\hpr P$ is initialized in $B_p$
(i.e.~$\vec x\in B_p$) then $\hpr P$'s state can reach, respectively,
will remain in $B_q$.  If $p$ is omitted, then $p$ is assumed to be
provided by the context.  We abuse $\bar p$, $p^\circ$, and
$\partial p$ to denote $B_p$'s complement, interior, and boundary.  We
use $=$ for definitions and equalities and $\equiv$ for abbreviations.

\subsection{Linearly Controlled, Damped, Plane Inverted Pendula}
\label{sec:ip-setting}

The inverted pendulum captures a range of applications (e.g.\ in
robotics) and is, thus, instructive for stability analysis.  
Consider a pendulum of mass~$m$ with a rigid rod of length~$l$ (the
distance from its center of rotation and its center of mass).  With
the angle $\theta$ between rod and vertical axis, the upper
equilibrium point is $\theta_e=0$.  The pendulum is subjected to
friction (proportional to its angular velocity~$\omega$ and a
constant~$f$) while being controlled for bringing it to and keeping it
at $\vec x_e=[\theta_e,\omega_e]^\intercal=\vec 0$.  The dynamics are given
by
\begin{align}
  \label{def:ip-dynamics-nonlin}
  \dot{\vec x}=
  \begin{bmatrix}
    \dot\theta\\\dot\omega
  \end{bmatrix}
  =
  \begin{bmatrix}
    \omega\\
    a\theta + b\omega + c\sin\theta
  \end{bmatrix}
\end{align}
where $a=\frac{k_1}{ml}$, $b=\frac{fl + k_2}{ml}$, and
$c=-\frac{g}{l}$ with gravity~$g$.

We assume there to be a linear controller of the form
$\vec u=\vec k^\intercal\vec x$ with tuning parameters
$\vec k = [k_1 k_2]^\intercal$ and corresponding to a tangential force
applied to the center of mass.  If $\vec u$ is generated by a momentum
around the center of rotation, we have $\vec k=\frac{\vec k_m}{l}$
where $\vec k_m$ contains the parameters of a torque controller
$\vec u_m$.
The discussion below is, thus, largely independent of whether $\vec k$
comes from linear proportional-derivative %
or linear-quadratic regulator %
design and identification, how $\vec k$ is interpreted, or whether
$\vec k$ is chosen without considering performance criteria more
specific than asymptotic stability.

\section{Characterizing Stable Inverted Pendula}
\label{sec:stability}
\label{sec:ip-lyap-family}

We illustrate, by example of the pendulum, (i) stability assurance of
non-linear systems using Lyapunov's direct method and, inspired by
\cite{Khalil2015-Nonlinearcontrol}, (ii) the characterization of
stable problem families by deriving constraints on the parameters of a
given dynamics~$\dot{\vec x}$ and an appropriate Lyapunov function
template~$V$.

It is well-known \cite{Khalil2015-Nonlinearcontrol} that quadratic
forms $\vec x^\intercal\vec P\vec x$ can be used to modify a system's
potential and kinetic energy function to meet necessary criteria for
definiteness.  Based on typical energy modeling of inverted pendula,
we then obtain
\begin{align}
  \label{def:ip-lyap}
  V &=\frac{ml^2}{2}\big(p_{11}\theta^2 + 2p_{12}\theta\omega
      + p_{22}\omega^2 - 2c(1-\cos\theta)\big)
\end{align}
where $\vec P=\left[p_{ij}\right]_{i,j\in 1,2}$ is a symmetric matrix
to be defined in a way to establish \eqref{constr:lst-nd} and
\eqref{constr:lst-pd}.  This $V$ is an example of a Lyapunov function
template that characterizes the stable family of pendula of the type
described by \eqref{def:ip-dynamics-nonlin}.

From matrix analysis, we know that \eqref{constr:lst-nd} can be
achieved by $p_{11}p_{22}-p_{12}^2>0$ and $p_{11}>0$.  For
\eqref{constr:lst-pd}, \eqref{def:ip-lyap} and
\eqref{def:ip-dynamics-nonlin} yield
\begin{align}
  \label{eq:ip-lyap-deriv}
  \dot V = \frac{ml^2}{2}\vec x^\intercal\mat Q\vec x
  + ml\big((g-gp_{22})\omega - gp_{12}\theta\big)\sin\theta
\end{align}
with symmetric 
\begin{align*}
  \mat Q =
  \begin{bmatrix}
    2ap_{12}
    &p_{11}+bp_{12} + ap_{22}
    \\p_{11}+bp_{12} + ap_{22}
    &2(p_{12}+bp_{22})
  \end{bmatrix}.
\end{align*}
From $mlg(1-\cos\theta)$ being positive definite, $f, g, l, m > 0$,
and $a, b, c<0$, the coefficients in the expansion of
\eqref{eq:ip-lyap-deriv} suggest a simplification by $p_{22}=1$.  These
observations allow us to under-approximate the stable family of
inverted pendula by the following constraints on $\dot{\vec x}$ (via
$a,b$) and $\mat P$:
\begin{align}
  \label{constr:lyap-func-k2}
  p_{11}&>p_{12}^2
  \;\land\; p_{12}>0 
  \;\land\; -\frac{p_{11}}{p_{12}}<b<-p_{12}
  \\\label{constr:lyap-class-eq}
  \text{and} \quad
  a&=-p_{11}-bp_{12}\,.
\end{align}
A proof using Lyapunov's direct method is possible if these
constraints are satisfiable.\footnote{Unsatisfiable constraints deny
  conclusions about the stability of $\dot{\vec x}$.  Because of the
  under-approximation, there might still be some $V$ that could work.
  In particularly, definiteness analysis of $-\mat Q$ not further
  pursed here may allow constraint relaxation and lead to a more
  complete method.}  $\mat P$ can then be chosen such that
sign-indefinite terms in \eqref{eq:ip-lyap-deriv} are canceled.
Depending on $a,b$, \eqref{constr:lyap-class-eq} determines, with
$p_{11},p_{12}$, the family of stable pendula.

Resolving \eqref{constr:lyap-class-eq} provides $p_{11}=-a-bp_{12}$.
With $p_{12}$ in \eqref{constr:lyap-func-k2}, we can express the
pendula family based on \eqref{def:ip-lyap} and
\eqref{eq:ip-lyap-deriv}.  For $V$ to remain positive definite, we
have to test $0<p_{12}<\sqrt{p_{11}}$ from above, the latter of which
is $p_{12}<\sqrt{-a-bp_{12}}$.

\paragraph*{Model Validation}

\begin{figure}[t]
  \includegraphics[width=\linewidth]{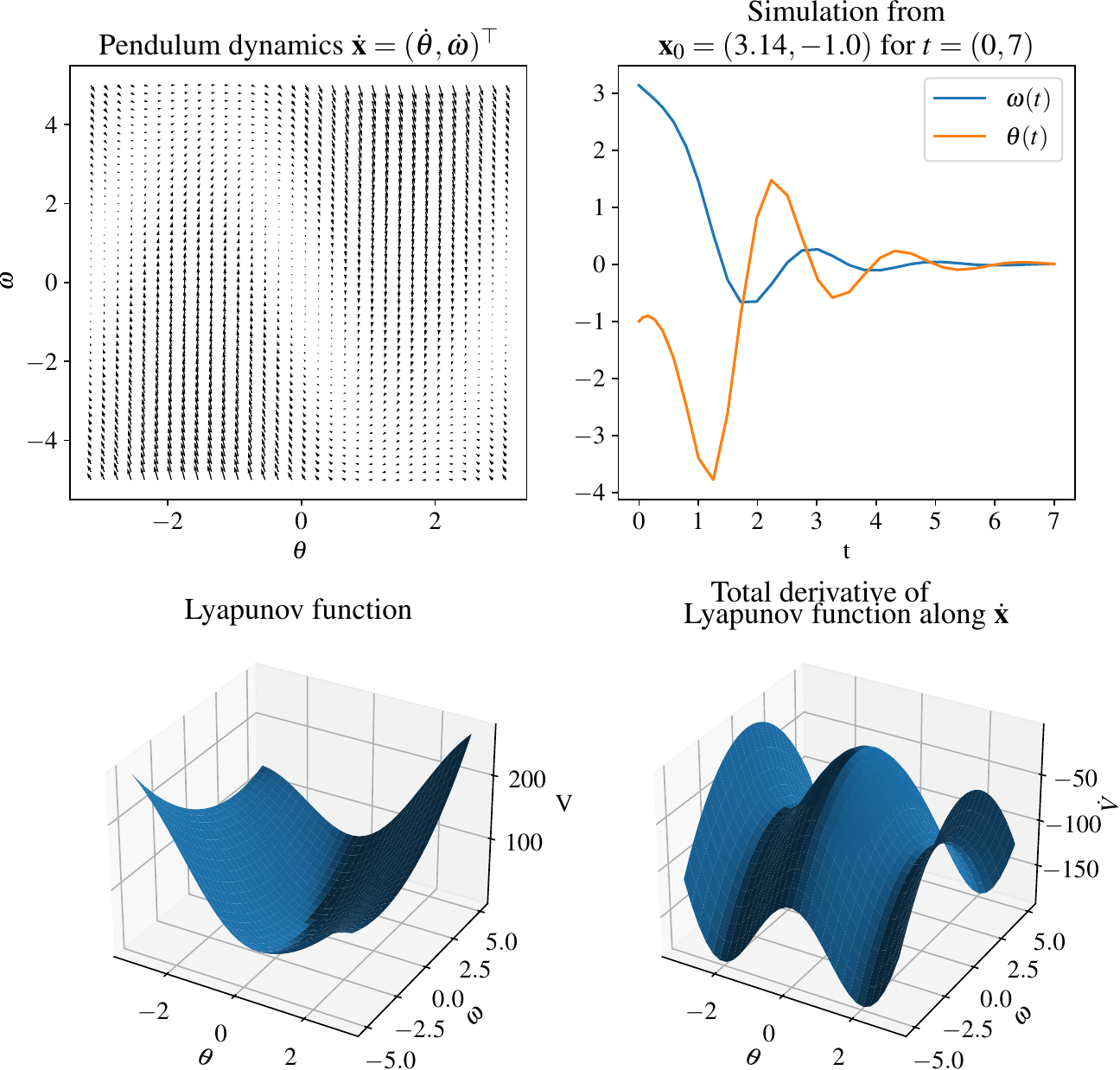}
  \caption{The trigonometric pendulum variant according to
    \eqref{def:ip-dynamics-nonlin}: the vector field of $\dot{\vec x}$
    (upper left), a simulation (upper right), $V$ (lower left), and
    its total derivative $\dot V$ (lower right)}
  \label{fig:ip-non-linear-valid}
\end{figure}

Our plausibility checks for the development of
\eqref{def:ip-dynamics-nonlin} and \eqref{def:ip-lyap} are informed by
simulation and an understanding of the dynamics (a vector field) and
the shape of $V$ and~$\dot V$ as developed in 
Fig.~\ref{fig:ip-non-linear-valid} (and~Fig.~\ref{fig:ip-linear-valid}).

\begin{figure*}
  \centering
  \includegraphics[width=.8\linewidth]{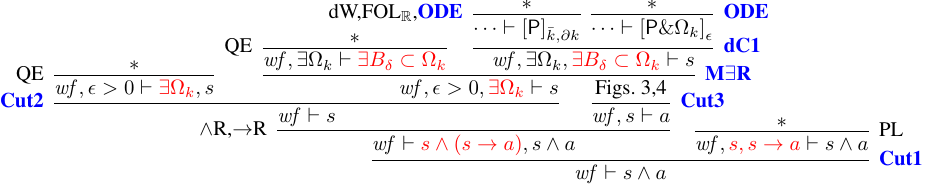}
  \caption{Part 1: Decomposing asymptotic stability into stability ($s$)
    and attractivity ($a$) in $\mathsf{d}\mathcal{L}$-sequent style extracted from
    \cite{DBLP:conf/tacas/TanP21}}
  \label{fig:generic-proof-1}  
  \label{fig:generic-proof}
\end{figure*}

\section{Deductive Proofs of Asymptotic Stability}
\label{sec:deductive-proof}

We now revisit the mechanization of a proof of Lyapunov's stability
theorem \cite{DBLP:conf/tacas/TanP21} in KeYmaera~X.  This
mechanization can be used for stability assurance of non-trigonometric
systems.  We add a proof parameter~$\mathit{wf}$ for well-formedness, which
can make use of \eqref{constr:lyap-func-k2} and
\eqref{constr:lyap-class-eq} and thereby reduce proof complexity while
maintaining sound reasoning.

Lyapunov's stability theorem enables the use of the direct method
(Sec.~\ref{sec:lst}).  Its proof \cite[Theorem
4.1]{Khalil2015-Nonlinearcontrol} uses an
$\epsilon$-$\delta$-construction.  It has been mechanized
\cite{DBLP:conf/tacas/TanP21} in $\mathsf{d}\mathcal{L}$\ (Sec.~\ref{sec:dl-intro}) and
applied to a non-trigonometric pendulum in Sec.~\ref{sec:ip-setting}.

\paragraph*{Running Example}

We adopt the method in \cite{DBLP:conf/tacas/TanP21} for proving a
similar variant where sine and cosine in
\eqref{def:ip-dynamics-nonlin} and \eqref{def:ip-lyap} are replaced
with their first- and second-order Taylor expansions
$\sin\theta\approx\theta$ and
$\cos\theta\approx 1 - \frac{1}{2}\theta^2$.  The reason for such
approximations are well-known limitations of quantifier
elimination~(QE) of trigonometric and general polynomials.  As a
result, we obtain a linear-quadratic problem comprising
\begin{align}
  \label{def:ip-dynamics-lin}
  \dot{\vec x}
  &=\begin{bmatrix}
    \dot\theta\\\dot\omega
  \end{bmatrix}
    =
    \begin{bmatrix}
      \omega\\d\theta + b\omega
    \end{bmatrix}
    \quad\text{with}\;d=a+c,
\end{align}
represented as the purely continuous hybrid program 
\[
  \mathsf{IP}\equiv \{ \dot\theta= \omega, \dot\omega = d\theta +
  b\omega\}
\]
together with a non-trigonometric Lyapunov function
\begin{align}
  \label{def:ip-lyap-quad}
  V &=\frac{ml^2}{2}\big(-(d+bp_{12})\theta^2
    + 2p_{12}\theta\omega + \omega^2\big)
\end{align}
such that the problem can be handled rather straightforwardly by the
chosen tooling, KeYmaera~X and the Wolfram~Engine.

\subsection{Key Elements of the Proof}

We parameterize the $\mathsf{d}\mathcal{L}$\ sequent-style proof in
\cite{DBLP:conf/tacas/TanP21} by a program~$\hpr{P}$, a Lyapunov
function template~$V$, and a well-formedness condition~$\mathit{wf}$.
In our example, we use $\hpr{P} \equiv \hpr{IP}$,
\[
  \mathit{wf}\equiv~\eqref{constr:lyap-func-k2}
  \land\bigwedge_{i\in\{g,l,m\}}i>0
  \land\bigwedge_{j\in\{a,b,c,d\}}j<0
\]
and $V\equiv\eqref{def:ip-lyap-quad}$, but will continue with the
parametric proof.

\begin{figure*}
  \centering
  \includegraphics[width=\linewidth]{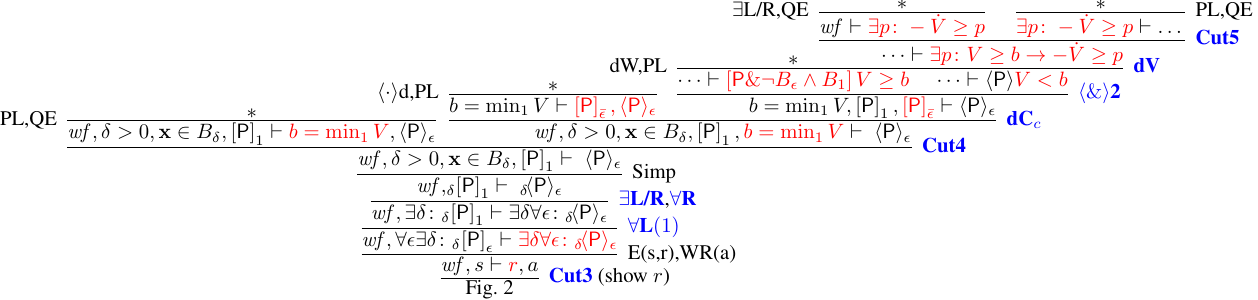}
  \caption{Part 2: Proving reachability ($r$) of the equilibrium point $\vec
    x_e$ and any of its environments}
  \label{fig:generic-proof-2}
\end{figure*}

The proof tree is split into three parts as illustrated in the
Figures~\ref{fig:generic-proof-2}, \ref{fig:generic-proof-1},
and~\ref{fig:generic-proof-3} reflecting its structure.
The tree is to be read from the root sequent at the bottom to the top.
If a node in the tree has two branches, we call the \emph{left}-hand
branch the \emph{show}-branch and the \emph{right}-hand branch the
\emph{use}-branch.

In part~1 (Fig.~\ref{fig:generic-proof-1}), the root sequent
$\mathit{wf} \vdash s \land a$ expresses asymptotic stability by concluding
stability ($s$) and attractivity ($a$), formally,
\begin{align*}
  s&\equiv \forall\epsilon\exists\delta\colon _\delta\dbox{\hpr{P}}_\epsilon
   \\&\equiv \forall\epsilon>0\exists\delta>0\forall \vec x\in B_\delta\colon
  \dbox{\hpr{P}} \vec x\in B_\epsilon
  \qquad\text{and}
  \\a&\equiv \exists\delta\forall\epsilon\colon
       _\delta\ddia{\hpr{P}}\dbox{\hpr{P}}_\epsilon
  \\&\equiv \exists\delta>0\forall \vec x\in
       B_\delta, \epsilon>0\colon
  \ddia{\hpr{P}}\dbox{\hpr{P}}\vec x\in B_\epsilon\,,
\end{align*}
from a well-formedness condition ($\mathit{wf}$) that system parameters (e.g.\
$a$,$b$,$c$) must satisfy. %

Condition $s$ specifies that $\hpr{P}$ remains within any $\epsilon$
environment if $\hpr{P}$ is initialized in the corresponding $\delta$
environment.  Condition $a$ holds if $\hpr{P}$ will reach and remain
in an arbitrarily small $B_\epsilon$ when starting from some~$B_\delta$.

Part~1 (Sec.~\ref{sec:proof-part-1}) splits the proof into a stability
proof and a proof that attractivity is implied by stability if the
origin $\vec x_e$ is stable \cite[Remark~3]{DBLP:conf/tacas/TanP21}.
One can, however, see that $s$ and $a$ do not generally imply each
other.  The stability proof follows the idea that within an arbitrary
$B_\epsilon$, one can find a largest set $\Omega_k\subseteq\mathbb{R}^n$
of states with $V\leq k$ (i.e.\ a $k$-levelset) enclosed in
$B_\epsilon$ and some~$B_\delta$ enclosed in $\Omega_k$.
\eqref{constr:lst-nd} and \eqref{constr:lst-pd} are used as
assumptions on the way to prove that any trajectory leaving
$B_\epsilon$ must have started outside $B_\delta$.
Moreover, Part~2 (Sec.~\ref{sec:proof-part-2}) handles the proving of
reachability of $\vec x_e$ from $\vec x_0$ by showing strictly
$V$-monotonic progress of $\hpr{P}$ towards~$\vec x_e$.
Finally, Part~3 (Sec.~\ref{sec:proof-part-3}) deals with proving
stability (in formal verification also called \emph{safety}) of
$\vec x_e$ based on reachability of~$\vec x_e$.

The proof tree explained below highlights key steps~(i.e.\ applying
inference rules to intermediate verification conditions in the
antecedents and succedents), particularly, cuts (\ref{eq:cut}, i.e.\
deductive shortcuts using auxiliary conditions and requiring user
interaction).  For brevity, we omit steps non-essential for
understanding the structure (e.g.\ weaken, unfolding) and occasionally
hide conditions in the antecedents and succedents if they are
irrelevant for the current step.  The most relevant rules are
explained in Sec.~\ref{sec:dl-intro} and
App.~\ref{sec:further-proof-rules}.

\subsection{Part 1: Decomposing Asymptotic Stability}
\label{sec:proof-part-1}

The proof of the main theorem (the root sequent in
Fig.~\ref{fig:generic-proof-1}) comprises several key steps shown
separately below.

\dLrule{Cut1}: Observed by \cite[Remark~3,
Cor.~5]{DBLP:conf/tacas/TanP21}, the proof of asymptotic stability
gets simpler if $\vec x_e$ is stable.  We, hence, introduce the
premise $s \land (s \to a)$ of modus ponens as an auxiliary
condition.  This cut allows us to separately prove $s$, $s\to a$, and
via modus ponens, $s\land a$.

\dLrule{Cut2}: Given $\epsilon>0$, we search for the largest
$k$-levelset
\[
  \Omega_k=\{\vec x\in\mathbb{R}^n\mid V(\vec x)\leq k\}
\]
enclosed in ball $B_\epsilon$.  Because of $V(\vec x(0))\geq 0$ and
$\dot V(\vec x(t))\leq 0$ (derived from assumptions
\ref{constr:lst-nd} and \ref{constr:lst-pd}), any trajectory $\xi$
starting in $\Omega_k$ remains in $\Omega_k$.  We use the cut
condition
\[
  \exists\Omega_k
  \equiv
  \exists k>0\forall\vec x\colon\norm{\vec x}=\epsilon\to V(\vec
  x)\geq k
\]
defining this $k$.  For the non-trigonometric pendulum,
$\exists\Omega_k$ can be checked automatically from
$\mathit{wf}\land\epsilon>0$ by QE.

\dLrule{M$\exists$R}: To prove $s$ from $\exists\Omega_k$, we now look
for a ball $B_\delta$ with $0<\delta<\epsilon$ inside $\Omega_k$ such
that $\forall\vec x\in B_\delta\colon V(\vec x)<k$.  To facilitate
this, we cut the condition $\exists B_k\subset\Omega_k$ into the proof
using the specific cut rule M$\exists$R.

\dLrule{dC1}: Next, by a differential cut of $\hpr{P}$, the conditions
introduced by Cut2 and M$\exists$R can now be used to perform two
sub-proofs: First, that whenever $\hpr{P}$ is at boundary
$\partial B_\epsilon$ then it is either outside $\Omega_k$ or at
$\partial\Omega_k$, formally,
\[
  \dbox{\hpr{P}}(\vec x\in\partial B_\epsilon
  \to\vec x\in\bar\Omega_k\cup\partial\Omega_k)
\]
(for tree brevity, $\dbox{\hpr{P}}_{\bar k,\partial k}$) and,
second, that $\hpr{P}$, restricted to $\Omega_k$, remains inside
$B_\epsilon$ (indicated with
$\dbox{\hpr{P}\&\Omega_k}_\epsilon$).

\dLrule{ODE}: We resolve the unrestricted box modality
$\dbox{\hpr{P}}_{\bar k}$ using differential weaken (dW) and
first-order sequent inference (FOL$_\mathbb{R}$) and
$\dbox{\hpr{P}}_{\partial k}$ by ordinary differential equation (ODE)
solving.  The restricted box modality
$\dbox{\hpr{P}\&\Omega_k}_\epsilon$ on the right can be tackled
directly with automated ODE solving.

The ``*'' leaf in each branch of the tree signfies that
an axiom (i.e. a statement whose truth is justified outside $\mathsf{d}\mathcal{L}$) has
been reached.  As shown before, relating the top-most (i.e. most
detailed) proof steps with axioms is made easier by automated
inference, for example, QE and ODE solving.

\subsection{Part 2: Proving Reachability of the Equilibrium
  Point}
\label{sec:proof-part-2}

A key task in this part (Fig.~\ref{fig:generic-proof-2}) is to derive
$r$ from~$s$ by establishing a differential variant~(\dLrule{dV}),
representing the strictly monotonic decrease of $V$ along the
trajectories of $\hpr{P}$.

\dLrule{$\forall$},\dLrule{$\exists$}: We eliminate some quantifiers
first.  With \dLrule{$\forall$L$(1)$}, we fix the assumption
$\epsilon=1$, that is, for some $\delta$, $\hpr{P}$ can remain within
the unit circle $B_1$.  Then, \dLrule{$\exists$L} eliminates $\exists$
in the antecedent.  After $\exists$-elimination in the succedent with
\dLrule{$\exists$R}, $\delta$ is provided by the antecedent.

\dLrule{Cut4}: After simplifications (Simp), we cut in the
condition
\[
  (b=\min_1 V) \equiv \exists b \forall\vec x\in B_1\colon
  V(\vec x)\geq b\,,
\]
which characterises $V$'s minimum $b$ inside $B_1$.  By QE, we
verify whether such a $b$, consistent with $\mathit{wf}$, exists.

\dLrule{dC$_c$}: In the use-branch of Cut4, we introduce an auxiliary
condition $\dbox{\hpr{P}}_{\bar\epsilon}$ (i.e.\ $\hpr{P}$ remains
outside $B_\epsilon$), leading to the tautology
$\ddia{\hpr{P}}_\epsilon\lor\dbox{\hpr{P}}_{\bar\epsilon}$ in the
show-branch of dC$_c$.  In its use-branch,
$\dbox{\hpr{P}}_1\land \dbox{\hpr{P}}_{\bar\epsilon}\land b=\min_1 V$
is at our disposal: $\hpr{P}$ never leaves $B_1$, never reaches
$B_\epsilon$, and $B_1$ contains the $b$-levelset.

\dLrule{$\ddia{\&}$2}: We show (i) using auxiliary condition
$V(\vec x)\geq b$\footnote{Here, we do not need to make use of the
  assumption $V(\vec x_e)=0$.} on the left for any $B_\epsilon$, that
trajectories inside $B_1$ and outside $B_\epsilon$ remain outside the
$b$-levelset, and (ii) with an auxiliary condition $V(\vec x)<b$ on
the right that there is a trajectory reaching the $b$-levelset.  We
have to show that (i) and (ii) can only be true simultaneously if
$\epsilon=0$, meaning that $\vec x_e$ is attractive for $\hpr{P}$.
($\dbox{\hpr{P}}_{\bar\epsilon}$ can only be true for $\epsilon=0$,
hence, $B_\epsilon$ for an arbitrarily small $\epsilon>0$ must be
reachable.)  While (i) can be shown from the antecedent by dW and
propositional reasoning (PL), (ii) needs a reachability proof.

\dLrule{dV}: The reachability proof can be tackled by proving a
differential invariant that amounts to checking that $\dot V<0$.  The
dV rule rephrases this check into
\begin{align*}
  \label{eq:dv}
  \exists p>0\forall\vec x\in \bar{B}_\epsilon\cap B_1\colon
  V(\vec x)\geq b \to -\dot V(\vec x)\geq p
\end{align*}
that is, checking for monotonic progress $-\dot V\geq p>0$ of $V$
along $\hpr{P}$'s trajectories in $B_1\cap \bar B_\epsilon$ and
apart from minimum~$b$.

\dLrule{Cut5}: Inside $B_1$, $b=\min_1 V$ implies $V\geq b$.  So, we
perform a cut with
\[
  (\exists p\colon-{\dot V}\geq p)
  \equiv
  \exists p>0\forall\vec x\in \bar{B}_\epsilon\cap B_1\colon
  -\dot V(\vec x)\geq p
\]
to (i) check the differential variant via QE in the context of $\mathit{wf}$
in the antecedent and (ii) use it for proving
$\ddia{\hpr{P}}_\epsilon$.

\begin{figure*}
  \centering
  \includegraphics[width=.7\linewidth]{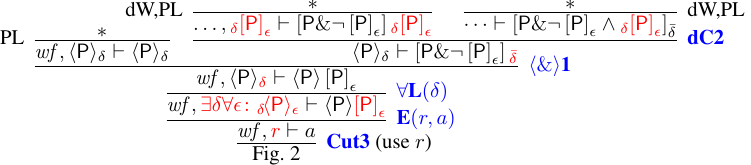}
  \caption{Part~3: Proving stability of the equilibrium point $\vec x_e$
    based on its reachability}
  \label{fig:generic-proof-3}
\end{figure*}

\subsection{Part 3: Proving Stability of the Equilibrium Point}
\label{sec:proof-part-3}

In Fig.~\ref{fig:generic-proof-3}, we continue with the proof of
$s\to a$, the observed implication between stability and attractivity
under the condition of a stable $\vec x_e$.  We can, only then,
establish attractivity as a special case of stability.

\dLrule{Cut3}: Attractivity requires that from $B_\delta$ (to be
proven to exist in the stability proof), an arbitrarily small
$B_\epsilon$ can be reached, formally,
\[
  r\equiv
  \exists\delta\forall\epsilon\colon_\delta\!\ddia{\hpr{P}}_\epsilon\equiv
  \exists\delta>0\forall\vec x\in B_\delta,\epsilon>0\colon
  \ddia{\hpr{P}}\vec x\in B_\epsilon.
\]
This auxiliary condition is introduced by Cut3.  Unlike where
M$\exists$R is applied, $B_\delta$ can now also include $B_\epsilon$.

\dLrule{E$(r,a)$}: $r$ and $a$ are merely expanded here.  Note that
asymptotic behavior implies that once $\hpr{P}$ reaches $B_\epsilon$
from $B_\delta$, it stays there.  Hence, note
the $\dbox{\hpr{P}}_\epsilon$ in the succedent.

\dLrule{$\forall$L($\delta)$}: After eliminating $\exists\delta$, we
fix $\epsilon=\delta$ in the antecedent to connect the reachability
condition~$r$ with the $\ddia{\cdot}$-modality of the attractivity
condition~$a$, eliminating $\forall\epsilon$ on the left and reducing
our focus to the proof of $\ddia{\hpr{P}}_\delta$, that is, that we
can reach an arbitrarily small $B_\delta\subseteq B_\epsilon$ for any
$\epsilon$ in the succedent.

\dLrule{$\ddia{\&}$1}: We decompose
$\ddia{\hpr{P}}\dbox{\hpr{P}}_\epsilon$ into a $\delta$-reaching
$\ddia{\cdot}$-modality (the show-branch establishes the
aforementioned connection) and an $\epsilon$-safe
$\dbox{\cdot}$-modality.  The use-branch of $\ddia{\&}1$ produces the
contra-positive of $\ddia{\hpr{P}}\dbox{\hpr{P}}_\epsilon$, namely,
$\dbox{\hpr{P}\&\neg\dbox{\hpr{P}}_\epsilon}_{\bar\delta}$.  That is,
evolution in states not $\epsilon$-invariant must occur outside
$B_\delta$, even outside $\Omega_k$.  (Because the
antecedent provides reachability of $B_\delta\subset\Omega_k$,
$\hpr{P}$ remains in $B_\epsilon$.)

\dLrule{dC2}: To the use-branch of $\ddia{\&}1$, we apply a
differential cut with the auxiliary condition
$_\delta\!\dbox{\hpr{P}}_\epsilon$, that is, $\epsilon$-safety for
some positive $\delta$, to be established as an invariant in the
show-branch.  Both dC2-branches can be closed by weakening (dW).  On
the left, we show that from $_\delta\dbox{\hpr{P}}_\epsilon$, formula
$\dbox{\hpr{P}\&\neg\dbox{\hpr{P}_\epsilon}}$ is vacuously true
(discharged by PL).  On the right, we then use the fact that there can
be no state in the intersection of
$\neg\dbox{\hpr{P}}_\epsilon\land{}_\delta\!\dbox{\hpr{P}}_\epsilon$,
concluding that $\hpr{P}$ must be outside $B_\delta$ in order to leave
an arbitrarily small $B_\epsilon$.

\section{Discussion}
\label{sec:discussion}

In the previous section, we have given an in-depth review of a deductive 
stability proof using the Lyapunuv method, which has been formalized in the 
Keymaera~X prover.
It has hopefully become clear that the level of detail at which such proofs are 
conducted goes beyond the level usually employed by mathematicians or 
engineers. 

An important benefit of working at this level, and of the mechanized proof 
check enabled by this effort, is that we can re-run the proof after changes, 
ensuring these changes do not invalidate the proof, and that we can be certain 
that all side conditions and assumptions are made explicit. 
For example, in the presented proof, the family of Lyapunov functions comes as 
a handy test for the associated controller design: we can change the 
controller, possibly plug in a changed Lyapunov function, and re-run the proof. 
We can be sure if our side conditions (e.g.\xspace constraints on the parameters) are 
too weak, the mechanized proof will likely fail. 

In Sec.~\ref{sec:stability}, we derived constraints on the problem parameters 
($a,b,\dots, \mat P$) to enrich the well-formedness condition~$\mathit{wf}$.  The 
strength of $\mathit{wf}$ can thus be better adjusted and, thus, aid both, the proof 
assistant and its user, in understanding and closing the relevant parts of a 
stability proof instance. 

Sec.~\ref{sec:deductive-proof} further indicates opportunities for new features 
on the prover side to make deductive systems and mechanized proof more readily 
available to the practising control engineer.  For example, a proof tree 
visualization (similar to that in Keymaera~X) is essential in understanding the 
proof structure.  In Keymaera~X, proofs are conducted and stored in the 
Bellerophone tactical language.  With more complex proofs requiring a range of 
cuts, it could, for example, be beneficial to navigate several tree branches 
side-by-side in addition to navigating at the Bellerophone level.

The presentation of proofs at an appropriate level, as illustrated in 
Sec.~\ref{sec:deductive-proof} for the stability proof, is crucial in following 
the successive application of proof rules, including their role in completing 
the proof. This allows proof authors and users (e.g.\ researchers, 
practitioners, certification auditors) to validate whether a proven theorem 
faithfully represents the  application problem, for example, a stability 
assurance problem.  Particularly important for these roles will be the validity 
of the assumptions included in the well-formedness condition as well as the 
approach to identifying auxiliary conditions for the various forms of cuts as 
applied in the example proof in Sec.~\ref{sec:deductive-proof}.  An 
understanding of both these aspects will support the ability of engineers to 
re-run, modify, and transfer existing proofs into different settings.

\section{Conclusion}
\label{sec:conclusion}

Deduction systems in general and proof assistants in particular facilitate 
obtaining stronger formal, detailed and mechanized mathematical proofs; 
that are more qualified to serve as guarantees of safety and other critical 
properties. 
In this paper, we address one of the main barriers hindering wider adoption of 
deduction systems in control engineering research and practice; 
namely the understandablity of the proofs and the complexity of the derivation 
steps. 
We explained how proof assistants can map manual proofs into mechanized ones by	
carrying on two stability proofs: a control-theoretic mathematical proof, and a 
one produced using the automated deduction tool Keymaera~X. 
The steps to guide the deduction system to complete the same proof mechanically 
was explained, and we highlighted the insights gained from doing the proof 
using the proof assistant Keymaera~X. 
The steps presented serve as a guide for how to carry on such a proof using a 
deduction system. 
This guidance can enable wider use and reuse of the proofs and deduction 
systems by control engineers,  in ways that contribute to more safe and 
formally verified controllers. 

\bibliographystyle{plain} %
\bibliography{paper}

\appendix
\section{Appendix}

\subsection{Further Proof Rules}
\label{sec:further-proof-rules}

Let $\hpr{C}\equiv\dot{\vec x}=f(\vec x)$ be a purely continuous
hybrid program in the following.  For reference, we provide an
overview of further $\mathsf{d}\mathcal{L}$\ sequent inference rules used in this work:
\begin{itemize}
\item[\dLrule{dC$_c$}] The compatible ODE cut rule, defined as
  \[
    \infer[\dLrule{dC$_c$}]
    {\Gamma \vdash C(\ddia{\hpr{C}\&Q}P), \Delta}
    {\Gamma, \dbox{\hpr{C}\&Q}R
      \vdash C(\ddia{\hpr{C}\&Q}P), \Delta
      & \Gamma \vdash \dbox{\hpr{C}\&Q}R, \Delta}
  \]
  can prepare for proving a differential variant (\dLrule{dV}).  The
  branches in \dLrule{dC$_c$}'s premise enable showing and using the
  compatible box modality assumption $\dbox{\hpr{C}\&Q}R$.
  
\item[\dLrule{$\ddia{\&}$}] The domain diamond rule, defined as
  \[
    \infer[\dLrule{$\ddia{\&}$}]
    {\Gamma \vdash \ddia{\hpr{C}\&Q}P, \Delta}
    {\Gamma \vdash \ddia{\hpr{C}\&Q}R, \Delta
     & \Gamma \vdash \dbox{\hpr{C}\&Q\land\neg P}\neg R, \Delta}
  \]
  can prepare for proving a differential variant (\dLrule{dV}) based
  on postcondition $R$ implying the original condition~$P$.  The
  contrapositive $\neg P\to\neg R$ thereof is encoded in
  \dLrule{$\ddia{\&}$}'s right-hand branch.
  
\item[\dLrule{dW}] The differential weaken rule, defined as
  \[
    \infer[\dLrule{dW}]
    {\Gamma\vdash\dbox{\hpr{C}\&Q}P,\Delta}
    {\Gamma_{const},Q\vdash P,\Delta_{const}}
  \]
  allows one to use the circumstance that if $Q$ holds and $\hpr{C}$
  is restricted to $Q$, $P$ holds independent of $\hpr{C}$'s
  execution.
\end{itemize}

\subsection{Supplemental Materials}

\begin{figure}
  \includegraphics[width=\linewidth]{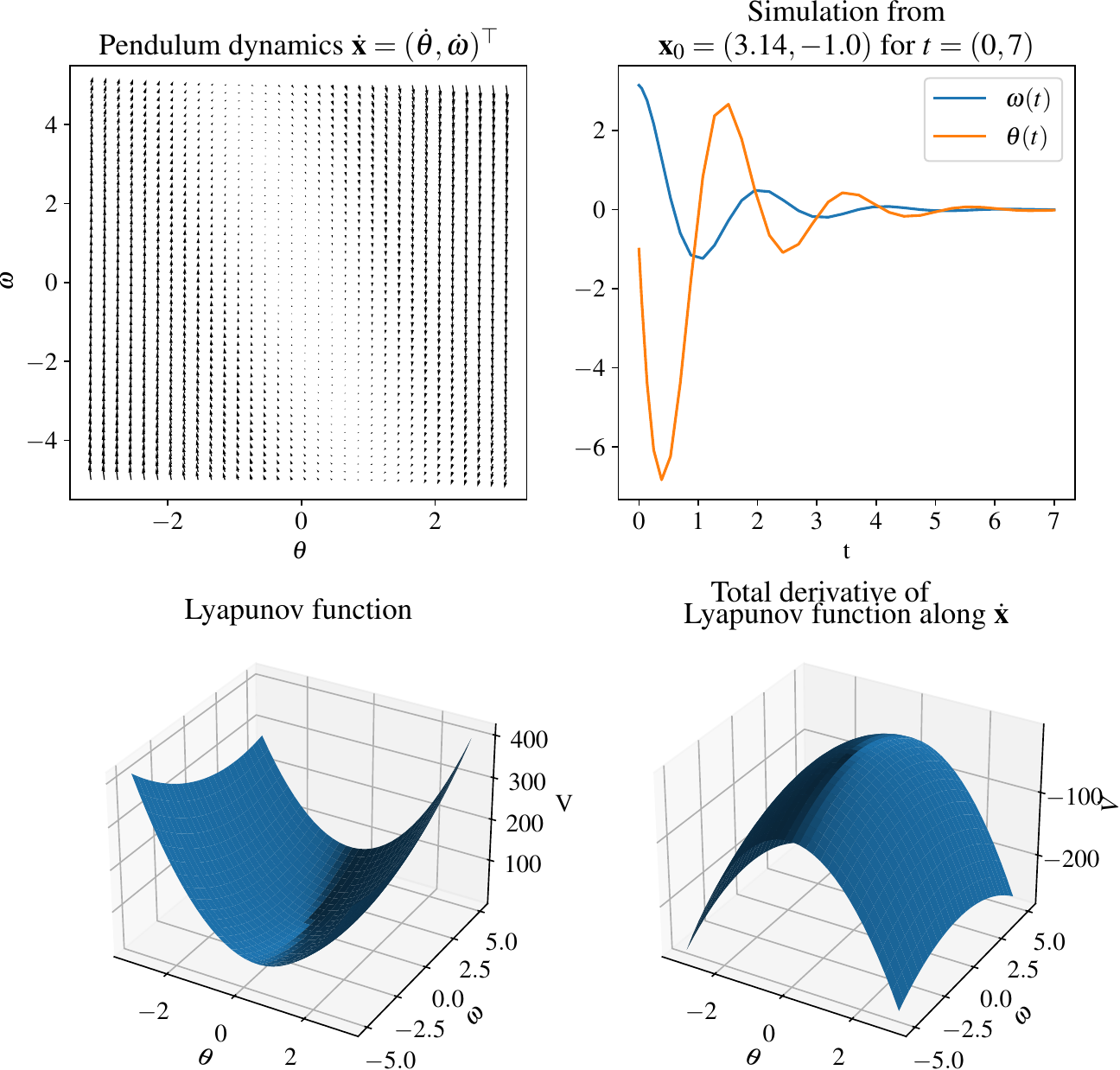}
  \caption{The non-trigonometric variant according to
    \eqref{def:ip-dynamics-lin}}
  \label{fig:ip-linear-valid}
\end{figure}

For comparison, Fig.~\ref{fig:ip-linear-valid} provides the validation
data set for the non-trigonometric pendulum controller.

\begin{figure}
  \centering
  \includegraphics[width=\linewidth]{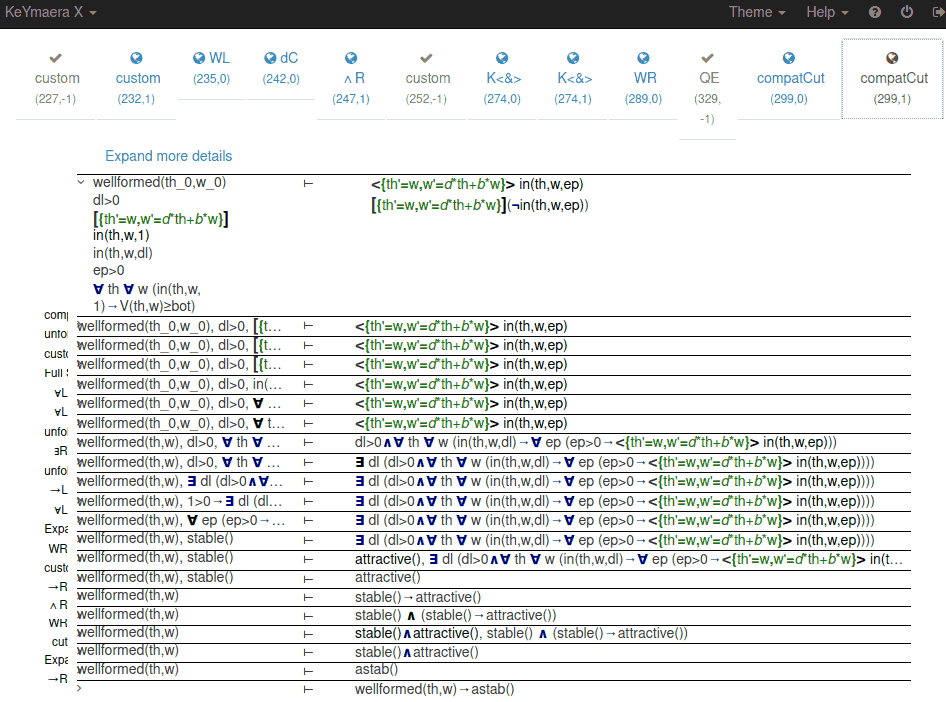}
  \caption{Navigating the finalized $\mathsf{d}\mathcal{L}$\ sequent proof in the
    KeYmaera~X tool}
  \label{fig:keymaera-screenshot}
\end{figure}

Fig.~\ref{fig:keymaera-screenshot} shows the user interface of the
KeYmaera~X proof assistant for navigating through the branches of a
finalized $\mathsf{d}\mathcal{L}$\ sequent proof tree.

\end{document}